\documentclass[preprint,12pt]{elsarticle}
\usepackage{amsmath}
\usepackage{amssymb}
\usepackage{setspace}
\usepackage[dvipsnames,svgnames]{xcolor}
\usepackage{tikz}
\usepackage{pgflibraryarrows}
\usepackage{pgflibrarysnakes}

\journal{arXiv}

\begin{document}

\begin{frontmatter}

\title{Competing spreading processes and immunization in multiplex networks}

\author[label1]{Bo Gao}\ead{gaobonmghhht@gmail.com}
\author[label2]{Dawei Zhao}\ead{zhaodw@sdas.org}
\address[label1]{School of Computer Information management, Inner Mongolia University of Finance and Economics, Hohhot 010051, China}
\address[label2]{Shandong Provincial Key Laboratory of Computer Networks,
Shandong Computer Science Center (National Supercomputer Center in
Jinan), Jinan 250014, China}

\begin{abstract}
Epidemic spreading on physical contact network will naturally introduce the human awareness information diffusion on virtual contact network, and the awareness diffusion will in turn depress the epidemic spreading, thus forming the competing spreading processes of epidemic and awareness in a multiplex networks. In this paper, we study the competing dynamics of epidemic and awareness, both of which follow the $SIR$ process, in a two-layer networks based on microscopic Markov chain approach and numerical simulations. We find that strong capacities of awareness diffusion and self-protection of individuals could lead to a much higher epidemic threshold and a smaller outbreak size. However, the self-awareness of individuals has no obvious effect on the epidemic threshold and outbreak size. In addition, the immunization of the physical contact network under the interplay between of epidemic and awareness spreading is also investigated. The targeted immunization is found performs much better than random immunization, and the awareness diffusion could reduce the immunization threshold for both type of random and targeted immunization significantly.
\end{abstract}


\end{frontmatter}


\section{Introduction}

In the past years, complex network approach has proven to be a successful tool in describing a large variety of real-world complex systems, ranging from biological, technological, social to information, engineering, and physical systems \cite{newman2010networks,barrat2008dynamical,cohen2010complex,newman2003structure,strogatz2001exploring,boccaletti2006complex,albert2000error,gao2013principle,gao2013attractor}. However, most of previous works are mainly concentrated to the case of single network which treats all the network's links on an equivalent footing \cite{newman2010networks,barrat2008dynamical,cohen2010complex}. Such network modeling methods may occasionally result in not fully capturing the details present in some real-life problems, leading even to incorrect descriptions of some phenomena that are taking place on real-world systems. Recently, with the development of human cognition and ``big data", the focus on complex networks has been extended from single network to multiplex network which is composed of several network layers constructed by same nodes but with different topologies and dynamics \cite{gomez2013diffusion,gomez2012evolution,zhao2014multiple,sole2014centrality,min2015link,sola2013eigenvector,rossi2015towards,boccaletti2014structure}. Multiplex network explicitly captures the authentic and natural characteristics of real world systems: the same node may have different kinds of interactions and each channel of connectivity is represented by a layer. Thus far, the topological and dynamical characteristics of multiplex networks and various of dynamical process (such as epidemic spreading \cite{zhao2014multiple,min2013layer,liu2016community,buono2014epidemics,zuzek2015epidemic}, evolutionary game \cite{nakamura2015evolutionary,perc2013collective,matamalas2015strategical,wang2015evolutionary,di2015quantifying,gomez2012evolution} and synchronization \cite{gambuzza2015intra,bogojeska2013opinion,yuan2014exact,dwivedi2015optimization}) upon them have attracted great attention in both theoretical and empirical areas, and a lot of remarkable results have been achieved.

As one of the hottest research topics of complex network science, epidemic spreading dynamic has centered on the modeling of different type of spreading processes and their control strategies \cite{newman2010networks,barrat2008dynamical,pastor2015epidemic,pastor2001epidemic,moreno2004dynamics,gao2011network,gao2011network,chen2008finding,cohen2003efficient,pastor2002immunization,madar2004immunization,zhao2013efficient}. The most successful epidemiology models include susceptible-infected ($SI$) model, susceptible-infected-susceptible ($SIS$) model, and susceptible-infected-recovered ($SIR$) model, both of which are good proxies for many real spreading processes
involving disease in human contact networks,
information and rumor in social networks, and virus in
computer or communication networks, etc \cite{newman2010networks,barrat2008dynamical,pastor2015epidemic,pastor2001epidemic,moreno2004dynamics}. Correspondingly, many mitigation and prevention strategies of epidemics are also proposed, one of the most popular
and effective methods is network immunization, such as random immunization, targeted immunization and acquaintance immunization, etc. \cite{gao2011network,gao2011network,chen2008finding,cohen2003efficient,pastor2002immunization,madar2004immunization}, where certain nodes in a network acquire immunity, and are thus no longer able to transmit the disease to their neighbors.

With the advent of multiplex networks, the traditional epidemic models and control methods
were extended to incorporate the structure of multiplex networks. The most interesting topics are the multiple routes spreading processes \cite{zhao2014multiple,min2013layer,liu2016community,buono2014epidemics,zuzek2015epidemic}, and their immunization \cite{zuzek2015epidemic,zhao2014immunization,wang2015immunity,buono2015immunization}. In addition, another rapidly evolving research, the competing spreading on multiplex networks, has recently
attracted considerable attentions \cite{granell2013dynamical,granell2014competing,wei2016unified,wang2014asymmetrically,alvarez2016competing,sahneh2014competitive,wei2016cooperative,fan2016effect,massaro2014epidemic,wang2016suppressing}. The most representative example is that disease spreading on physical contact network will naturally introduce the human awareness information diffusion on virtual contact network, and the awareness diffusion will in turn depress the epidemic spreading, thus forming the competing spreading processes of epidemic and awareness in a two-layers networks. Granell et al. study the dynamical interplay between epidemic and awareness, both of which follow the $SIS$ models, in multiplex networks. They found the critical onsets of both dynamics get intertwined and the onset of the epidemic starts depending on the incidence of aware individuals \cite{granell2013dynamical,granell2014competing}. Wang et al. also investigate these two type of spreading dynamics where the disease obeys the $SIRV$ model and the awareness the $SIR$ model, and find epidemic outbreak on the contact layer can induce an outbreak on the communication layer, and information spreading can effectively raise the epidemic threshold \cite{wang2014asymmetrically}.

In this paper, we study the competing dynamics of epidemic and awareness, both of which follow the $SIR$ process and the self-protection and the self-awareness of individual are also incorporated, in a two-layer networks based on microscopic Markov chain approach and numerical simulations. We will investigate the impacts of awareness diffusion and the capacities of self-protection and self-awareness of individuals on the epidemic threshold and the final outbreak size of the epidemic. Furthermore, the efficiency of random and targeted immunizations of multiplex network under the interplay between of epidemic and awareness spreading will be studied.

\section{Models and Analysis}

\begin{figure}[!htb]
\centering
\includegraphics[scale=0.5,trim=50 0 50 0]{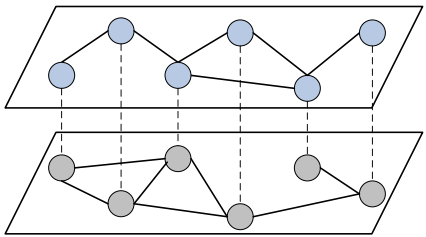}
\caption{A multiplex networks composed of two network layers
interrelated with each other, nodes are the same in both layers and
the connectivity inter-layer is from each node to itself.}\label{network}
\end{figure}

The proposed model consists
of a multiplex networks coupled by two network layers and two spreading processes
proliferated by each layer. As shown in Fig. \ref{network}, the up layer and below layer indicate the virtual contact network and physical contact network respectively, denoted by $A$ and $B$. Both of them have the same $N$ nodes with different intra-layer topologies. $(a_{ij})_{N\times N}$ and $(b_{ij})_{N\times N}$ are defined as the adjacency matrices of $A$ and $B$ respectively, where $a_{ij}=1$ indicates there is a link form node $i$ to node $j$ in layer $A$, otherwise $a_{ij}=0$, and a similar definition applies to $b_{ij}$.

For the spreading processes of awareness and epidemic, we assume both of them follow the $SIR$ epidemiology models. In the $SIR$ model, each node can be in one of the three states: susceptible state ($S$) in which the individual is free of the epidemic but can be infected via contacts with infected individuals; infected state ($I$), where the individual carries the disease and can transmit it to susceptible individuals; and recovered state ($R$), in which the individuals recovered from the disease and cannot pass the disease to other nodes or be infected again. The
classic SIR model uses discrete time step for its evolution and at
each time step, the infected node can infect its susceptible
neighbors with transmissibility $\beta$, and then becomes
recovered or removed node with probability $\delta$. Here, we denote $\beta_A$ ($\beta_B$) and $\delta_A$ ($\delta_B$) as the transmissibility and recover rate of the nodes in layer $A$ ($B$). Moreover, we assume the $R$ state nodes in layer $A$ still have the knowledge of risk information, but just have no willing to pass the information.

In our model, the awareness diffusion in layer $A$ and the epidemic spreading in layer $B$ are not two irrelevant processes, they are dynamic interplay and influence with each other: a node that is aware ($I$ state) in layer $A$ will take measures for preventing infection which is called the self-protection of the individual, this behavior can be reflected by the reduction of individual's
own infectivity with a factor $\gamma$ ($0 \leqslant\gamma \leqslant 1$) in layer $B$; a node that is infected in layer $B$ will become
aware in layer $A$ with probability $\kappa$ ($0 \leqslant\kappa \leqslant 1$), which indicates the self-awareness ability of individual due to the infection of the epidemic.

\begin{figure}[!htb]
\centering
\includegraphics[scale=0.5,trim=50 0 50 0]{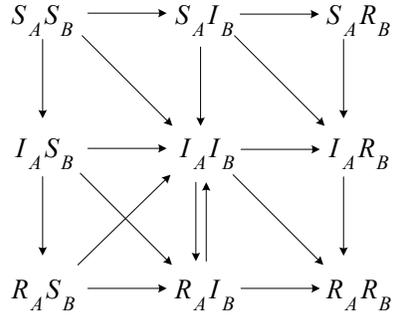}
\caption{Transitions between states of nodes, the arrow out from a given state of node at time step $t$ points to its possible successor state at time step $t+1$..}\label{transition}
\end{figure}

Summing up, in our proposed model, every node of the multiplex network falls into the following nine states: $S_AS_B$, $S_AI_B$, $S_AR_B$, $I_AS_B$, $I_AI_B$, $I_AR_B$, $R_AS_B$, $R_AI_B$ and $R_AR_B$, where $X_AY_B$ refers to node is $X$ state in layer $A$ and $Y$ state in layer $B$ respectively. Fig. \ref{transition} shows the possible transitions between states of nodes, the arrow out from a given state of node at time step $t$ points to its possible successor state at time step $t+1$. The transition probability $p_i^{X_AY_B\rightarrow X'_AY'_B}(t)$ from state $X_AY_B$ to its successor $X_AY_B$ of node $i$ at time step $t$ is given as follows:

\begin{equation}
\begin{split}
&p_i^{S_AS_B\rightarrow S_AI_B}(t)=q_i(t)(1-q_i^{S_A}(t))(1-\kappa),\\
&p_i^{S_AS_B\rightarrow I_AS_B}(t)=(1-q_i(t))q_i^{I_A}(t),\\
&p_i^{S_AS_B\rightarrow I_AI_B}(t)=q_i(t)(1-q_i^{S_A}(t))\kappa +(1-q_i(t))(1-q_i^{I_A}(t)),\\
&p_i^{S_AI_B\rightarrow S_AR_B}(t)=q_i(t)(1-\kappa)\delta_B,\\
&p_i^{S_AI_B\rightarrow I_AI_B}(t)=[1-q_i(1-\kappa)](1-\delta_B),\\
&p_i^{S_AI_B\rightarrow I_AR_B}(t)=[1-q_i(t)(1-\kappa)]\delta_B,\\
&p_i^{S_AI_B\rightarrow I_AR_B}(t)=[1-q_i(t)(1-\kappa)]\delta_B,\\
&p_i^{S_AR_B\rightarrow I_AR_B}(t)=1-q_i(t),\\
&p_i^{I_AS_B\rightarrow I_AI_B}(t)=(1-\delta_A)(1-q_i^{I_A}(t))+\delta_A(1-q_i^{I_A}(t))\kappa,\\
&p_i^{I_AS_B\rightarrow R_AS_B}(t)=\delta_Aq_i^{I_A}(t),\\
&p_i^{I_AS_B\rightarrow R_AI_B}(t)=\delta_A(1-q_i^{I_A}(t))(1-\kappa),\\
&p_i^{I_AI_B\rightarrow I_AR_B}(t)=(1-\delta_A)\delta_B,\\
&p_i^{I_AI_B\rightarrow R_AI_B}(t)=\delta_A(1-\delta_B)(1-\kappa),\\
&p_i^{I_AI_B\rightarrow R_AR_B}(t)=\delta_A\delta_B,\\
&p_i^{I_AR_B\rightarrow R_AR_B}(t)=\delta_A,\\
&p_i^{R_AS_B\rightarrow I_AI_B}(t)=(1-q_i^{I_A}(t))\kappa,\\
&p_i^{R_AS_B\rightarrow R_AI_B}(t)=(1-q_i^{I_A}(t))(1-\kappa),\\
&p_i^{R_AI_B\rightarrow I_AI_B}(t)=(1-\delta_B)\kappa,\\
&p_i^{R_AI_B\rightarrow R_AR_B}(t)=\delta_B,
 \end{split}
\end{equation}
where $q_i(t)$, $q_i^{S_A}(t)$ and $q_i^{I_A}(t)$ indicate the probabilities of node $i$ at time step $t$ not being informed by any neighbors, not being infected by any neighbors if $i$ was unaware, and not being infected by any neighbors if $i$ was aware, respectively. They are given by

\begin{equation}
\begin{split}
&q_i(t)=\prod_{j=1}^{N}(1-a_{ij}p_j^{I_A}(t)\beta_A),\\
&q_i^{S_A}(t)=\prod_{j=1}^{N}(1-b_{ij}p_j^{I_B}(t)\beta_B),\\
&q_i^{I_A}(t)=\prod_{j=1}^{N}(1-b_{ij}p_j^{I_B}(t)\gamma\beta_B),
 \end{split}
\end{equation}
where $p_j^{I_Z}(t)$ refers to the probability of node $j$ is $I$ state at time step $t$ in layer $Z$. Therefore, if we use $p_i^{X_AY_B}(t)$ denotes the probability of node $i$ is $X$ state in layer $A$ and $Y$ state in layer $B$ at time step $t$, we have
\begin{equation}
\begin{split}
&p_i^{I_A}(t)=p_i^{I_AS_B}(t)+p_i^{I_AI_B}(t)+p_i^{I_AR_B}(t),\\
&p_i^{I_B}(t)=p_i^{S_AI_B}(t)+p_i^{I_AI_B}(t)+p_i^{R_AI_B}(t).
 \end{split}
\end{equation}

 Based on above statements, the evolution of our proposed model can be expressed by the microscopic Markov chain approach equations which read as

\begin{equation}\label{Eq}
\begin{split}
&p_i^{S_AS_B}(t+1)=p_i^{S_AS_B}(t)(1-p_i^{S_AS_B\rightarrow S_AI_B}(t)-p_i^{S_AS_B\rightarrow I_AS_B}(t)-p_i^{S_AS_B\rightarrow I_AI_B}(t)),\\
&p_i^{S_AI_B}(t+1)=p_i^{S_AS_B}(t)p_i^{S_AS_B\rightarrow S_AI_B}(t)+p_i^{S_AI_B}(t)(1-p_i^{S_AI_B\rightarrow S_AR_B}(t)-\\
&\ \ \ \ \ \ \ \ \ \ \ \ \ \ \ \ \ \ \ p_i^{S_AI_B\rightarrow I_AI_B}(t)-p_i^{S_AI_B\rightarrow I_AR_B}(t)),\\
&p_i^{S_AR_B}(t+1)=p_i^{S_AI_B}(t)p_i^{S_AI_B\rightarrow S_AR_B}(t)+p_i^{S_AR_B}(t)(1-p_i^{S_AR_B\rightarrow I_AR_B}(t)),\\
&p_i^{I_AS_B}(t+1)=p_i^{S_AS_B}(t)p_i^{S_AS_B\rightarrow I_AS_B}(t)+p_i^{I_AS_B}(t)(1-p_i^{I_AS_B\rightarrow I_AI_B}(t)-\\
&\ \ \ \ \ \ \ \ \ \ \ \ \ \ \ \ \ \ \ p_i^{I_AS_B\rightarrow R_AS_B}(t)-p_i^{I_AS_B\rightarrow R_AI_B}(t)),\\
&p_i^{I_AI_B}(t+1)=p_i^{S_AS_B}(t)p_i^{S_AS_B\rightarrow I_AI_B}(t)+p_i^{S_AI_B}(t)p_i^{S_AI_B\rightarrow I_AI_B}(t)+\\
&\ \ \ \ \ \ \ \ \ \ \ \ \ \ \ \ \ \ \ p_i^{I_AS_B}(t)p_i^{I_AS_B\rightarrow I_AI_B}(t)+p_i^{R_AS_B}(t)p_i^{R_AS_B\rightarrow I_AI_B}(t)+\\
&\ \ \ \ \ \ \ \ \ \ \ \ \ \ \ \ \ \ \ p_i^{R_AI_B}(t)p_i^{R_AI_B\rightarrow I_AI_B}(t)+p_i^{I_AI_B}(t)(1-p_i^{I_AI_B\rightarrow I_AR_B}(t)-\\
&\ \ \ \ \ \ \ \ \ \ \ \ \ \ \ \ \ \ \ p_i^{I_AI_B\rightarrow R_AI_B}(t)-p_i^{I_AI_B\rightarrow R_AR_B}(t)),\\
&p_i^{I_AR_B}(t+1)=p_i^{S_AI_B}(t)p_i^{S_AI_B\rightarrow I_AR_B}(t)+p_i^{S_AR_B}(t)p_i^{S_AR_B\rightarrow I_AR_B}(t)+\\
&\ \ \ \ \ \ \ \ \ \ \ \ \ \ \ \ \ \ \ p_i^{I_AI_B}(t)p_i^{I_AI_B\rightarrow I_AR_B}(t)+p_i^{I_AR_B}(t)(1-p_i^{I_AR_B\rightarrow R_AR_B}(t)),\\
&p_i^{R_AS_B}(t+1)=p_i^{I_AS_B}(t)p_i^{I_AS_B\rightarrow R_AS_B}(t)+p_i^{R_AS_B}(t)(1-p_i^{R_AS_B\rightarrow I_AI_B}(t)-\\
&\ \ \ \ \ \ \ \ \ \ \ \ \ \ \ \ \ \ \ p_i^{R_AS_B\rightarrow R_AI_B}(t)),\\
&p_i^{R_AI_B}(t+1)=p_i^{I_AS_B}(t)p_i^{I_AS_B\rightarrow R_AI_B}(t)+p_i^{I_AI_B}(t)p_i^{I_AI_B\rightarrow R_AI_B}(t)+\\
&\ \ \ \ \ \ \ \ \ \ \ \ \ \ \ \ \ \ \ p_i^{R_AS_B}(t)p_i^{R_AS_B\rightarrow R_AI_B}(t)+p_i^{R_AI_B}(t)(1-p_i^{R_AI_B\rightarrow I_AI_B}(t)-\\
&\ \ \ \ \ \ \ \ \ \ \ \ \ \ \ \ \ \ \ p_i^{R_AI_B\rightarrow R_AR_B}(t)),\\
&p_i^{R_AR_B}(t+1)=p_i^{I_AI_B}(t)p_i^{I_AI_B\rightarrow R_AR_B}(t)+p_i^{I_AR_B}(t)p_i^{I_AR_B\rightarrow R_AR_B}(t)+\\
&\ \ \ \ \ \ \ \ \ \ \ \ \ \ \ \ \ \ \ p_i^{R_AI_B}(t)p_i^{R_AI_B\rightarrow R_AR_B}(t)+p_i^{R_AR_B}(t).
\end{split}
\end{equation}

\begin{figure}[!htb]
\centering
\includegraphics[scale=0.5,trim=50 0 50 0]{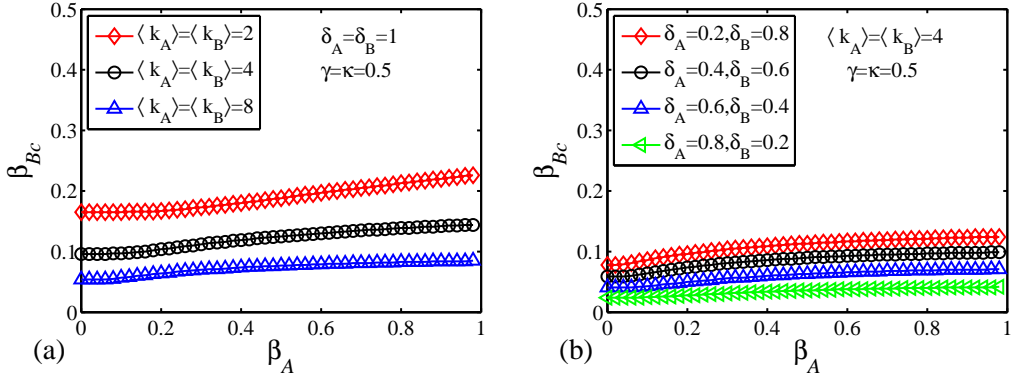}
\caption{The relationship between epidemic threshold ${\beta_B}_c$ and transmissibility $\beta_A$ of the awareness.}\label{iaib}
\end{figure}

\begin{figure}[!htb]
\centering
\includegraphics[scale=0.5,trim=50 0 50 0]{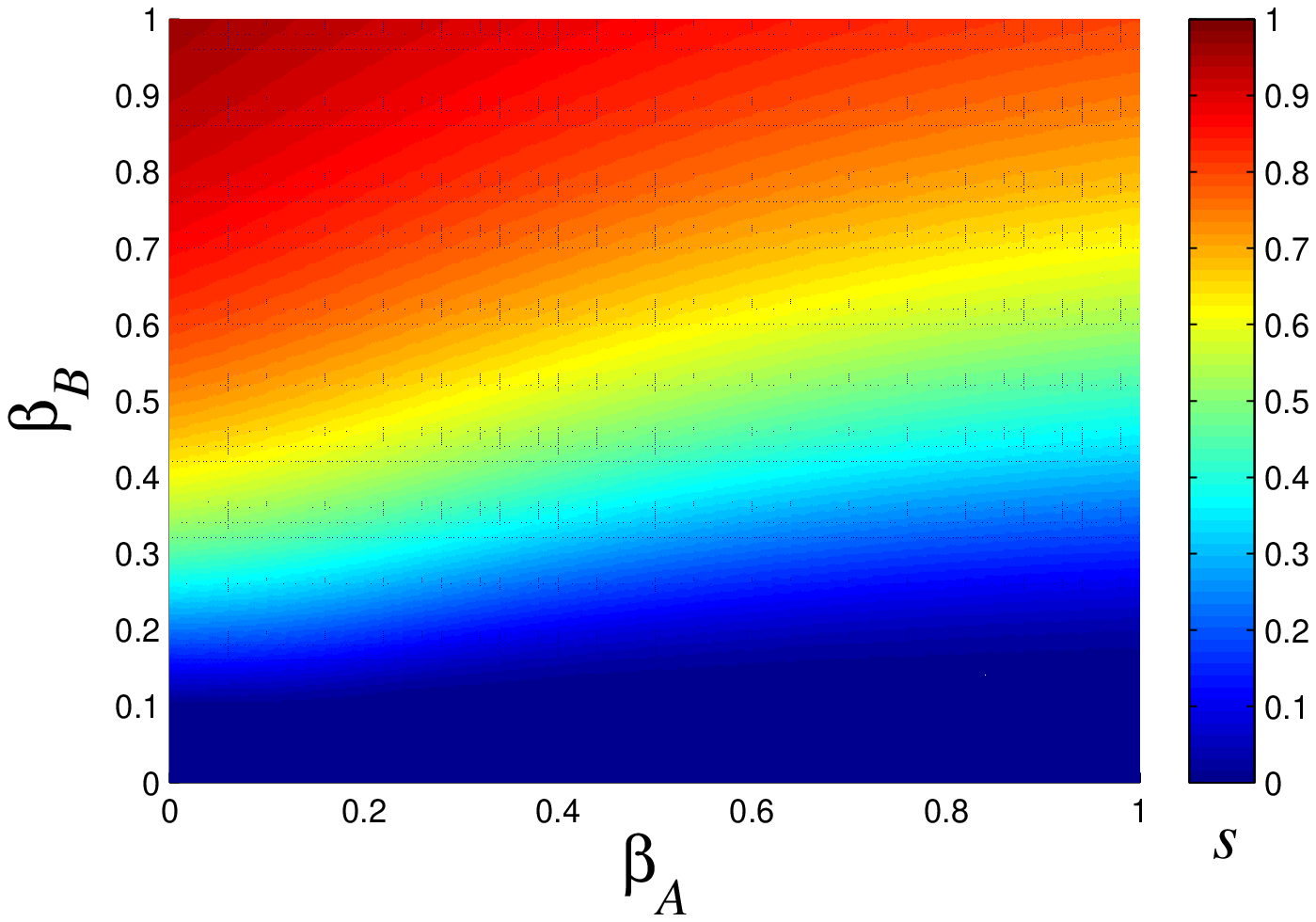}
\caption{The final outbreak size $s$ under different combinations of ${\beta_A}$ and $\beta_B$. The average degrees of the used networks are $\langle k_A\rangle=\langle k_B\rangle$=4, and $\delta_A$=$\delta_B$=1 and $\gamma$=$\kappa$=0.5.} \label{iaibsize}
\end{figure}

Due to the complicated interaction between the disease and
awareness spreading processes, the numerical calculation method is used to obtain the approximate threshold ${\beta_B}_c$ and outbreak size $s_B$ of the epidemic in layer $B$ based on Eq.4 which replaces the direct derivation. For all subsequent numerical simulations, we assume there is only one node carries the disease ($I$ state) in layer $B$ at the initial time stage which will be transferred in layer $B$  and introduces the awareness diffusion in layer $A$.
Fig.~\ref{iaib} features the relationship between the epidemic threshold ${\beta_B}_c$ and the transmissibility $\beta_A$ of the awareness. It is obviously found that ${\beta_B}_c$ increases with $\beta_A$ irrespective of the average degrees of multiplex networks (panel a) and the recover rates of the epidemic and the awareness (panel b). In addition, Fig.~\ref{iaibsize} gives values of final outbreak size $s_B$ of the disease under different combinations of ${\beta_A}$ and $\beta_B$. One sees that $s_B$ decreases with the increase of ${\beta_A}$. Both of Fig.~\ref{iaib} and ~\ref{iaibsize} indicate that the epidemic spreading on physical contact network induces the risk awareness diffusion on virtual contact network, and the awareness diffusion in turn depress the epidemic spreading.

\begin{figure}[!htb]
\centering
\includegraphics[scale=0.5,trim=50 0 50 0]{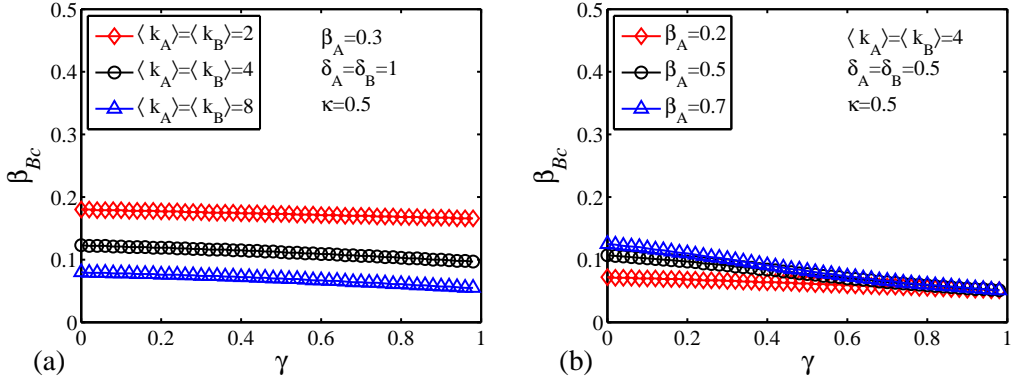}
\caption{The relationship between epidemic threshold ${\beta_B}_c$ and the capacity of self-protection $\gamma$ of individual.}\label{rib}
\end{figure}

\begin{figure}[!htb]
\centering
\includegraphics[scale=0.5,trim=50 0 50 0]{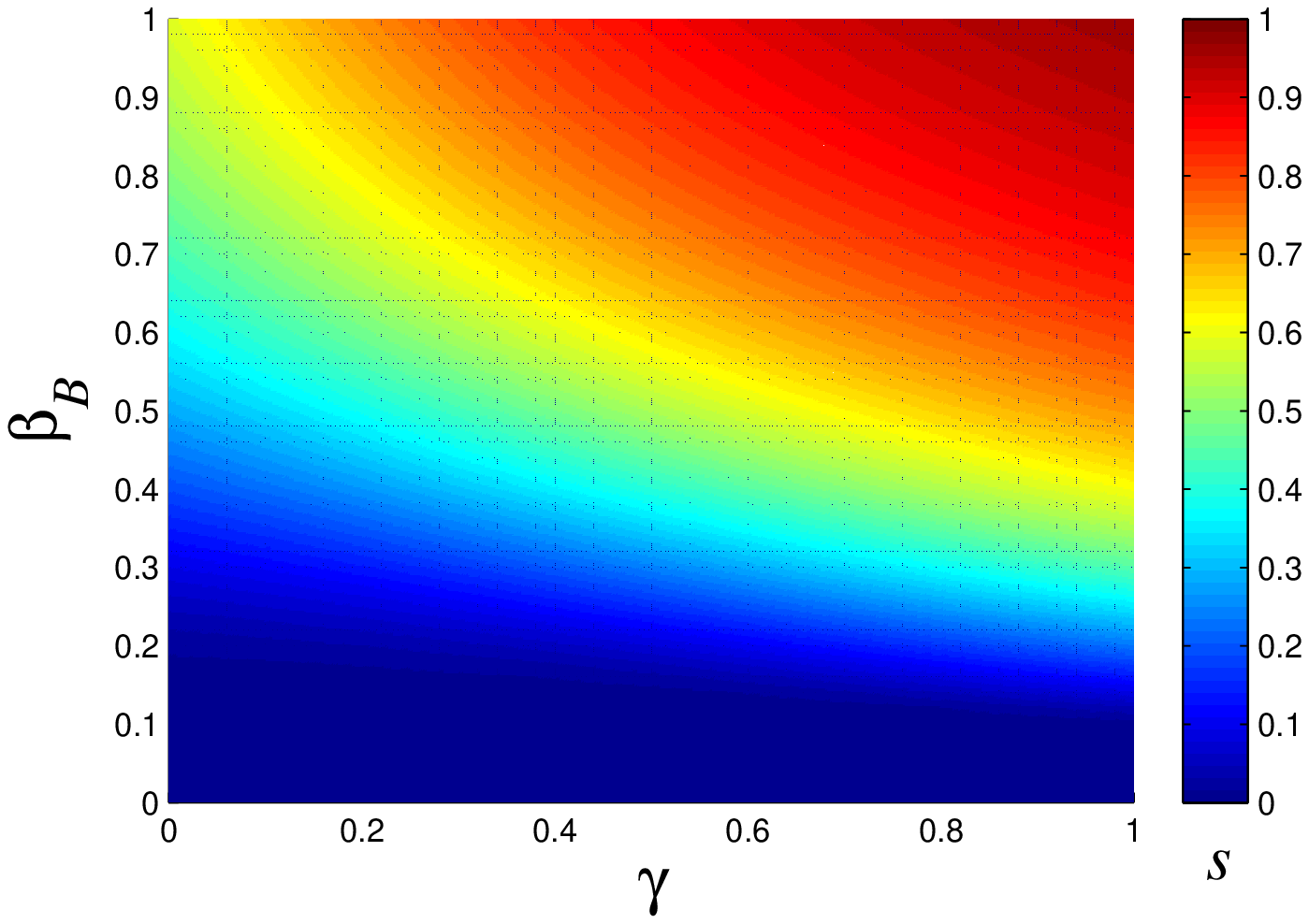}
\caption{The final outbreak size $s$ under different combinations of $\gamma$ and $\beta_B$. The average degrees of the used networks are $\langle k_A\rangle=\langle k_B\rangle$=4, and $\delta_A$=$\delta_B$=1 and $\beta_A$=$\kappa$=0.5.}\label{ribsize}
\end{figure}

\begin{figure}[!htb]
\centering
\includegraphics[scale=0.5,trim=50 0 50 0]{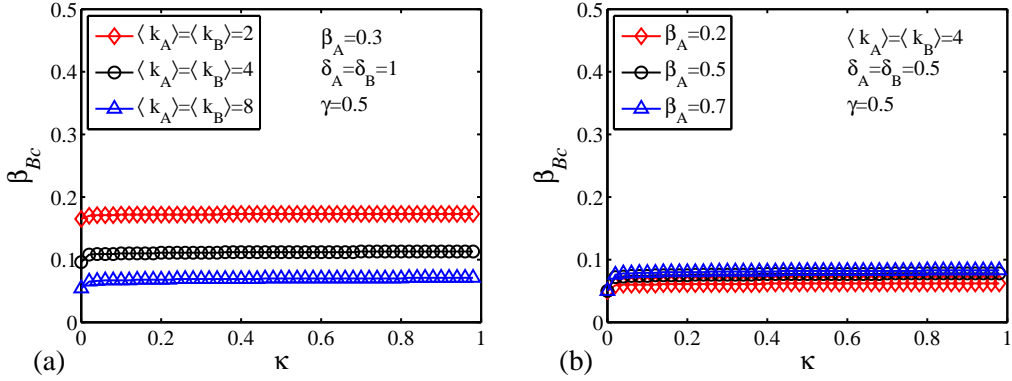}
\caption{The relationship between epidemic threshold ${\beta_B}_c$ and the capacity of self-awareness $\kappa$ of individual.}\label{kib}
\end{figure}

\begin{figure}[!htb]
\centering
\includegraphics[scale=0.5,trim=50 0 50 0]{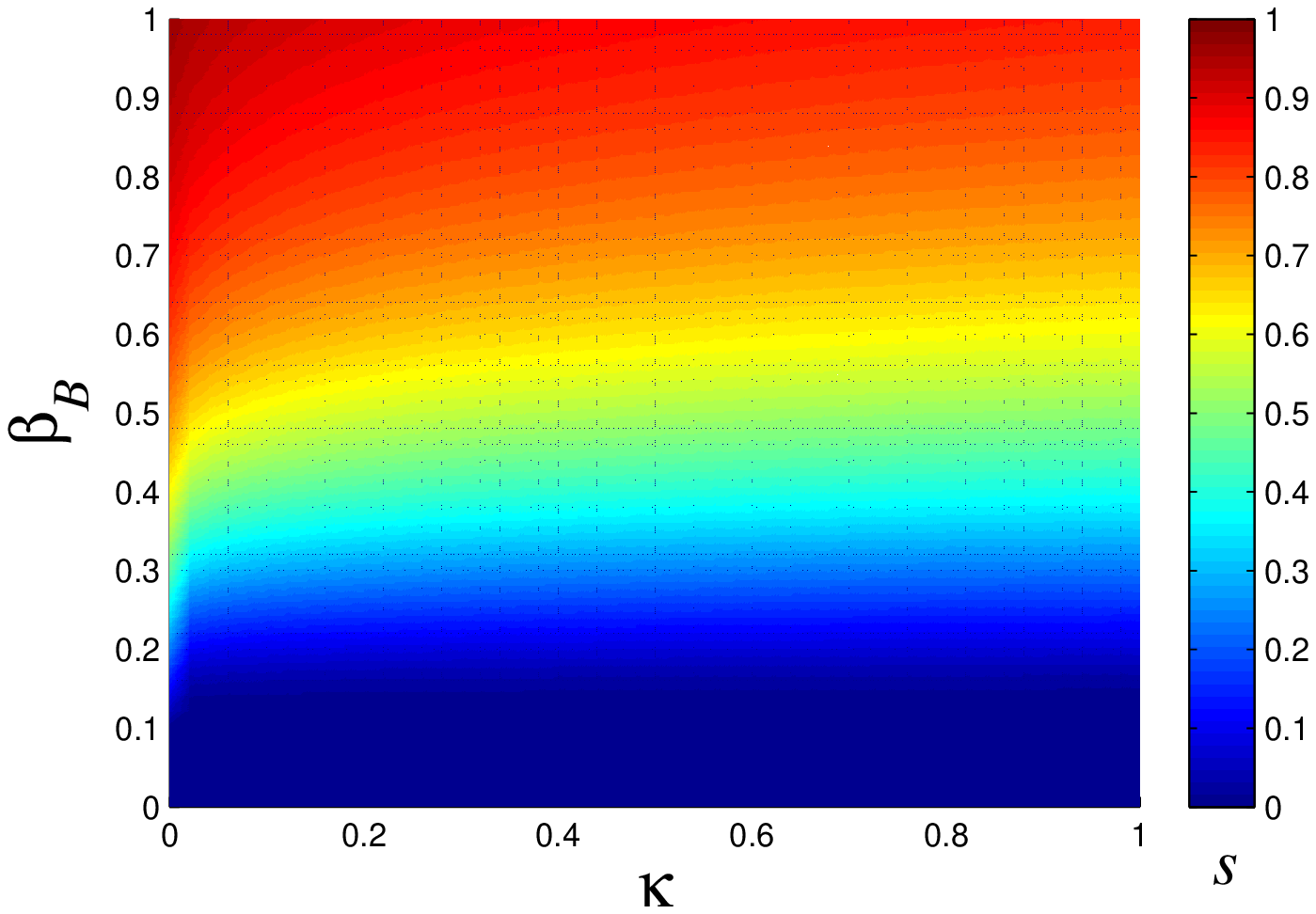}
\caption{The final outbreak size $s$ under different combinations of $\kappa$ and $\beta_B$. The average degrees of the used networks are $\langle k_A\rangle=\langle k_B\rangle$=4, and $\delta_A$=$\delta_B$=1 and $\beta_A$=$\gamma$=0.5.}\label{kibsize}
\end{figure}

The impacts of the capacities of self-protection $\gamma$ and self-awareness $\kappa$ of individuals on the epidemic threshold ${\beta_B}_c$ and the outbreak size $s_B$ are also investigated. From Fig.~\ref{rib}, one sees that the ${\beta_B}_c$ decreases with the increase of $\gamma$ irrespective of the average degrees of multiplex networks (panel a) and the transmissibility $\beta_A$ of the awareness (panel b). Since the smaller the $\gamma$, the stronger the capacities of self-protection of individuals are, and thus the larger the epidemic threshold will be. The values of final outbreak size of the disease under different combinations of ${\beta_B}$ and $\gamma$ is given in Fig.~\ref{ribsize}. It can be found that the outbreak size $s_B$ increases with both of the ${\beta_B}$ and $\gamma$. Moreover, we further uncover that ${\beta_B}$ and $s_B$ are not affected by the ability of the self-awareness $\kappa$ of individuals. As shown in Fig. \ref{kib} and Fig. \ref{kibsize}, we observe ${\beta_B}$ and $s_B$ are irrelevant to the values of $\kappa$ except the case of $\kappa=0$ which means no risk awareness diffusion in layer $A$. Together with the results of Fig.~\ref{iaib} and ~\ref{iaibsize}, we conclude that the strong capacities of awareness diffusion and self-protection of individuals could lead to a much higher epidemic threshold and a smaller outbreak size. However, the self-awareness of individuals has no obvious effect on the epidemic threshold and outbreak size.

\begin{figure}[!htb]
\centering
\includegraphics[scale=0.5,trim=50 0 50 0]{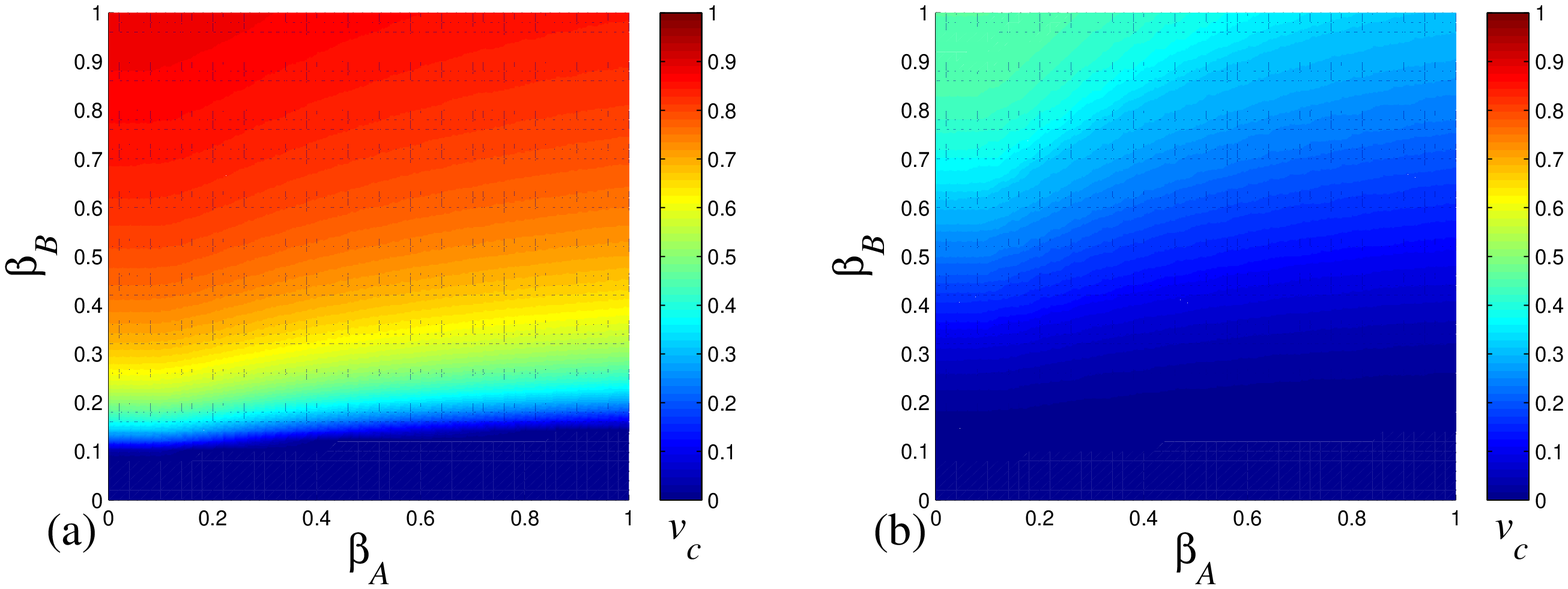}
\caption{The immunization threshold $v_c$ under different combinations of $\beta_A$ and $\beta_B$. (a) random immunization, (b) targeted immunization. The average degrees of the used networks are $\langle k_A\rangle=\langle k_B\rangle$=4, and  $\delta_A$=$\delta_B$=1 and $\gamma$=$\kappa$=0.5.}\label{iaibv}
\end{figure}

\section{Immunization and Analysis}

It can be found that, in the proposed model, the competing spreading of awareness and epidemic is free of external forces.    Therefore, if the risk awareness diffusion cannot depress the epidemic spreading completely, the other mitigation and prevention methods of epidemics are needed. To date, one of the most popular
and effective methods is network immunization, where certain nodes in network acquire
immunity, and are thus no longer able to transmit the disease to their neighbors. In this section, the random and targeted immunizations
of the physical contact network under the interplay between of epidemic and
awareness spreading are investigated. In this case, the $SIR$ epidemiology model takes place on layer $B$ is extended to the $SIRI'$ model in which the immunized state ($I'$) is added. Furthermore, the nodes of multiplex network have three more of new states including $S_AI'_B$, $I_AI'_B$  and $R_AI'_B$. The transitions between them are as follows
$$S_AI'_B \rightarrow I_AI'_B \rightarrow R_AI'_B.$$
However, these three states cannot transit to the other nine states mentioned in above section, and vice versa, since the immunizations are performed at the initial stage.

Based on above statements, the competing processes of awareness and epidemic incorporated with immunization can also be
expressed by the microscopic Markov chain approach equations, parts of which are given by

\begin{equation}\label{Eq}
\begin{split}
&p_i^{S_AI'_B}(t+1)=p_i^{S_AI'_B}(t)q_i(t),\\
&p_i^{I_AI'_B}(t+1)=p_i^{S_AI'_B}(t)(1-q_i(t))+p_i^{I_AI'_B}(t)(1-\delta_A),\\
&p_i^{R_AI'_B}(t+1)=p_i^{I_AI'_B}(t)\delta_A+p_i^{R_AI'_B}(t).\\
\end{split}
\end{equation}
And the other nine items including $p_i^{X_AY_B}(t)$ where $\{X,Y\}=\{S,I,R\}$, are defined as that in Eq.4, except the used parameter $p_i^{I_A}(t)$ becomes

$$p_i^{I_A}(t)=p_i^{I_AS_B}(t)+p_i^{I_AI_B}(t)+p_i^{I_AR_B}(t)+p_i^{I_AI'_B}(t).$$

In the initial stage of the model evaluation, $v$ fraction of nodes of layer $B$ are selected as immunized nodes. For the random immunization, the immunized nodes are selected uniformly at random. While for the targeted immunization, we focus on the degree-based strategy in which the immunized nodes are the first $v$ fraction of largest degree nodes of layer $B$.
Based on Eq. 5, we can calculate the threshold of random and targeted immunizations, above which the final outbreak size of the epidemic is null, via the numerical simulation. Fig. \ref{iaibv} show the result of immunization threshold $v_c$ under different combination of $\beta_A$ and $\beta_B$. It can be found that the threshold of targeted immunization (panel b) is much smaller than that of the random case (panel a) under same conditions, which means targeted immunization performs much better than random immunization for epidemic under the competing spreading of epidemic and awareness. We also find the threshold decreases with the increase of $\beta_A$ for both type of random and targeted immunizations, which indicates the awareness diffusion could reduce the immunization threshold effectively.
These results are enlightening in that the self protection of individual inspired by the risk awareness diffusion and the immunization from outside could help with each other to depress the epidemic spreading completely.


\section{Summary}

In this paper, we study the competing processes of epidemic spreading and awareness diffusion in a two-layer networks, and the capacities of the self-protection and self-awareness of individuals are also considered. An Markov chain functions are proposed to represent the evolution of the model, and numerical simulations are used to calculate the approximate epidemic threshold and the final outbreak size. We find the awareness diffusion and self-protection capacity of individuals could lead to a much higher epidemic threshold and a smaller outbreak size. However, the self-awareness of individuals has no obvious effect on the epidemic threshold and outbreak size. In addition, the immunization of the physical contact network under the interplay between of epidemic and awareness spreading is also investigated. The targeted immunization is found performs much better than random immunization, and the awareness diffusion could reduce the immunization threshold for both type of random and targeted immunization significantly.

\section{Acknowledgments} This paper was supported by the Inner Mongolia Colleges and Universities Scientific and Technological Research Projects (Grant no. NJZY132),the National Natural Science Foundation of China (No.31560622, No.31260538, No.30960246), the Shandong Province Outstanding Young Scientists Research Award Fund Project (Grant No. BS2015DX006) and the Shandong Academy of Sciences Youth Fund Project (Grant No. 2016QN003).




\bibliographystyle{unsrt}
\bibliography{bibtex}

\end{document}